*Comprehensive parameter and electrochemical dataset for a 1 Ah graphite/LNMO battery cell for physical modelling as a blueprint for data reporting in battery research*

**Authors**
Christina Schmitt, August Johansson, Xavier Raynaud, Eibar Joel Flores Cedeño, John Mugisa, Dane Sotta, Agathe Martin, Nicolas Schaeffer, Cédric Debruyne, Yvan Reynier, Simon Clark, Dennis Kopljar

**Affiliations**
1. Deutsches Zentrum für Luft- und Raumfahrt (DLR), Institute of Engineering Thermodynamics, 70569 Stuttgart, Germany
2. University of Stuttgart, Institute of Building Energetics, Thermal Engineering and Energy Storage (IGTE), 70569 Stuttgart, Germany
3. SINTEF Industry, Sustainable Energy Technology, Sem Sælands Vei 12, Trondheim, 7034, Norway
4. Univ. Grenoble Alpes, CEA, Liten, DEHT, 38000 Grenoble, France
5. SINTEF Digital, Mathematics and Cybernetics, PO Box 124 Blindern, 0314 Oslo, Norway

Corresponding author(s): Dennis Kopljar (dennis.kopljar@dlr.de)

**Abstract**
While current technology has enabled their widespread use, further improvements are needed for stationary, portable, and mobile applications, for example by the development of novel cathode materials. Digitalization of battery development, combining both experimental and modelling efforts is extremely valuable in this development. This is addressed in the present paper, where the authors present a comprehensive dataset for a graphite/LNMO 1 Ah pouch cell, including material, design, and electrochemical data. The dataset, validated through the BattMo modelling framework, supports physical modelling and aims to benefit the battery modelling community by offering a comprehensive resource for future studies. Both the dataset and the accompanying software for numerical validation is openly available and processed in such a way that it can serve as blueprint for reporting of comparable research data.

**Background & Summary**

Lithium-ion batteries are a key technology for the energy transition as they allow the integration of high shares of volatile renewable energy into the grid and enable the pivot from combustion engine cars to cleaner and more efficient electric vehicles. Although current technology has facilitated their introduction in an increasing share of applications, further development is still required to leverage the full potential for stationary, portable and mobile applications. The development process can be accelerated by a model-driven approach that combines experimental and modelling activities at an early stage. Although this approach is already state-of-the-art for cell and system development, its importance in the development process is continuously growing.

On their own, mathematical models can provide information about the coupling of quantities and allow correlating design choices to system trends. However, making accurate predictions about a specific cell or system requires high-quality parameters. On the one hand, equivalent circuit models (ECM) are computationally efficient and can be easily parameterized by pure electrochemical measurements such as pulses or impedance measurements at different SOC and temperature without knowledge about chemistry and physical parameters of the studied cell. Physics-based models on the other hand incorporate the actual physical and (electro-)chemical processes within the cell, thereby posing a more accurate representation of the cell. Their use is limited by the fact that they require a more extensive parameter set. This includes physicochemical parameters of the cell materials and components. The parameters contain information on the physics and the chemistry of the cell that together govern its macroscopic behavior and can generally be divided into five categories: design, structural, thermodynamic transport and kinetic parameters. Their extraction requires a range of different characterization methods and information from diverse sources.[1-5]

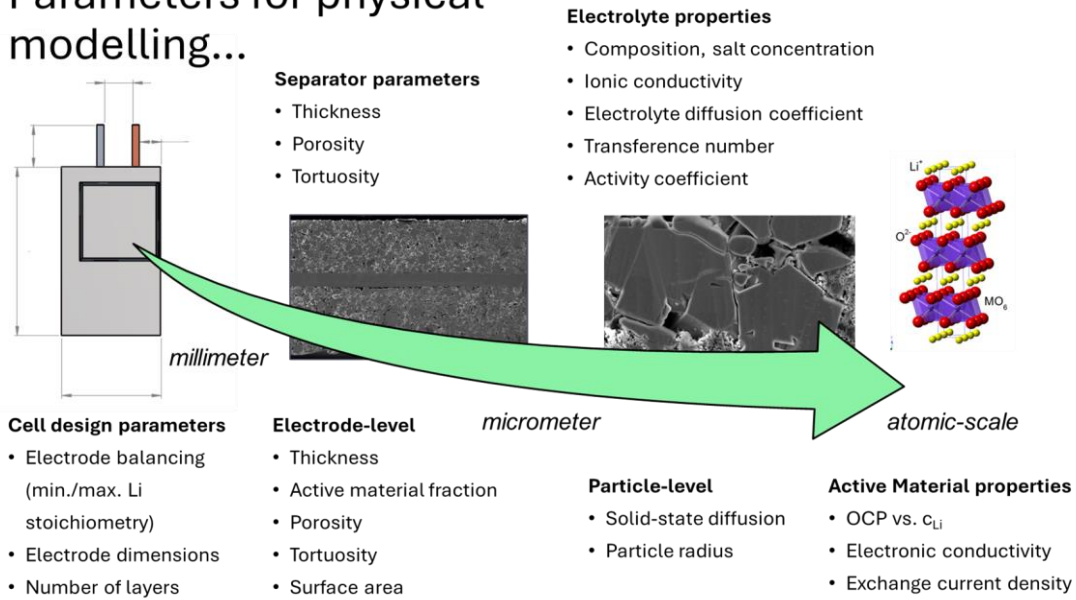

*Figure 1. Overview of parameters required for physical modelling and reported in this manuscript.*

Having access to these parameters is the precondition to model a specific cell with good accuracy. The procedure to extract the parameters involves the tear-down of the cell, very good experimental knowledge and access to a range of sophisticated equipment. Thus, it is very time and resource-consuming and only accessible to few facilities and institutions. The parameterization from input-output data via inverse modelling has gained traction lately and might mitigate this challenge to a certain degree, if challenges related to parameter identifiability are properly addressed [6].

Fortunately, an increasing number of scientific contributions have specifically worked on making such parameterization results for a range of commercial cells available in literature. Moreover, an open-source database called LiIonDB has been launched that collects parameter sets from literature and allows comparing certain parameter values between chemistries and publications.[7] Access to accurate parameter data is key to battery modelling and simulation; however, authors often describe the final parameters and the analysis procedure only. Raw data of the measurements from which the data was obtained, not to mention data rich in unambiguous meta-data is generally not included although it has tremendous value e.g. for the development of new algorithms for parameter extraction. This is an important research field as historically used approaches are often used without considering assumptions and simplifications that went into deriving model experiments and analysis methods. An important example is the use of GITT measurements to determine solid-state diffusion coefficient and the inconsiderate use of the standard analysis protocol derived in 1977 for planar, non-porous electrodes.[8] Furthermore, complete datasets on next-generation chemistry together with validation measurements of technical cells is very scarce although necessary to support the modelling of such advanced chemistries.

Motivated by that, we hereby publish the comprehensive parameterization results from a graphite/LNMO 1 Ah prototype pouch cell. The dataset includes data on materials, electrode and cell design as well as electrochemical data on electrode and pouch cell level. The data is complete for a physical modelling of the electrodes and cell, and is validated by using it as input for simulation in the opensource BattMo framework. Ultimately, the procedure that was followed for model calibration and the final outcome is described demonstrating the validity of the reported parameter set versus experimental discharge curves recorded at various C-rates for the described pouch cell design.

## Methods

### Electrode preparation

Negative electrode formulation was based on 96 wt.% graphite from Vianode, 1 wt.% Super C45 conductor additive (Imerys), 1.5 wt.% CMC (7HXF, Ashland) and 1.5 wt.% SBR latex (TRD105, JSR). The slurry was prepared in a double-walled water-cooled stainless-steel container using a high-shear disperser (Dispermat, VMA-Getzmann). First, C45 conductor additive was dispersed in a 2 wt.% CMC aqueous solution. Then, the water content was adjusted in the slurry to obtain an optimized viscosity. After that, graphite active material was weighed and introduced into the slurry before applying a high-speed dispersion step. Finally, SBR latex was added and the complete slurry was gently mixed for a few minutes. The final solid content of the slurry was 48 %. The positive electrode formulation was defined as follows: 92 wt.% LNMO (Johnson Matthey), 4 wt.% Super C65 conductor additive (Imerys) and 4 wt.% PVDF (Solef® 5130, Syensqo). The slurry was produced using a process similar to the negative electrode. First, C65 conductor additive was dispersed in a 8 wt.% PVDF solution in N-methylpyrrolidone (NMP) solvent. NMP was further added into the slurry to obtain adequate viscosity for dispersion, then LNMO active material was weighed and introduced into the slurry before dispersing at high speed. The final solid content of the slurry was 51 %.

Single-side (for electrochemical characterization) and double-side coated (for stack pouch cells) negative electrode were manufactured using a comma-bar coating machine (Ingecal) available in CEA pre-pilot line dry room (dew point -20 °C). The slurry was deposited on 10 µm copper or 20 µm aluminium current collector. Then, the electrodes passed through an inline dryer to remove any solvent trace. Finally, the electrodes were calendared at 80 °C to the target density using a two-roll calendaring equipment (Ingecal).

**Table 1. Electrode design parameters.**

|  | Units | Negative electrode | Positive electrode |
|---|---|---|---|
| Mass loading active material | mg/cm² | 7.3 | 17.8 |
| Active material content | wt.-% | 96 | 92 |
| Active material density | g/cm³ | 2.25 | 4.2 |
| Binder content | wt.-% | 1.5 (SBR) 1.5 (CMC) | 4 |
| Binder density | g/cm³ | 1.05 (SBR) 1.59 (CMC) | 1.8 |
| Conductive agent content | wt.-% | 1 | 4 |
| Conductive agent density | g/cm³ | 1.8 | 1.8 |
| Coating thickness | µm | 56.5 | 76.5 |
| Current collector thickness | - | 10 | 20 |
| Current collector type | | Cu | Al |
| Current collector density | g/cm³ | 8.96 | 2.7 |

**Cell assembly and electrochemical characterization**
Electrodes with 18 mm diameter were punched out of single-side coated electrodes and assembled in ECC-PAT-Core-Cells (EL-CELL) in a three-electrode configuration. Negative and positive electrodes were built versus a lithium metal counter electrode (500 µm, MSE Supplies) with a 260 µm thick Whatman GF/A separator and an integrated lithium reference ring (EL-CELL) to measure the working electrode versus the reference electrode (RE) potential (0 V versus Li/Li$^+$). The electrolyte used was 1.2 M LiPF$_6$ in a mixture of ethylene carbonate (EC) and ethyl methyl carbonate (EMC) (3:7 w:w) with 0.5 wt% fluoroethylene carbonate (FEC) supplied by Solvionic.

After assembly, the cells were transferred to a climate chamber (Memmert, IPP750) operating at 25 °C and tested with a BaSyTec cell test system (CTS Lab). Before characterization, the cells underwent formation cycles to form a stable SEI for material characterization according to the following sequence: 1 x C/10 and 1 x C/5 with CC-CV charge and CC discharge (CV-step until C/50 was reached). For the graphite cells, two C/5 cycles were applied to ensure a proper SEI formation on graphite. The voltage limits were 3 – 4.9 V for LNMO and 0.01 – 1.5 V in case of graphite. For the latter, the CV-step was applied at 0.01 V, which corresponds to the charge step in a full cell. In this work, the term "charge" always corresponds to the charge direction in a full cell. C-rates for charge and discharge were calculated based on the active material loading. The open circuit potentials (OCP) of the electrodes were measured by a quasi-OCP (qOCP) with C/50 and by galvanostatic intermittent titration technique (GITT). The GITT-OCP consisted of C/10 pulses with a duration of 2.5 min (corresponding to 240 points and 0.417% SOC intervals) followed by a relaxation phase until either ΔU < 0.5 mV/s for 30 min or t > 6 h.

In addition, the solid diffusion coefficient of lithium in LNMO and graphite is derived from the respective GITT measurements, based on the approach of Weppner and Huggins.[8] This approach posits that the voltage relaxation after a pulse results from an equilibration of Li$^+$ concentration gradients in the cell, and assumes that i) such equilibration is rate-limited by Li$^+$ diffusion within the solid particles of active material, and ii) the solid diffusion obeys the one-dimensional Fick's diffusion laws. Hence, solid diffusion coefficient *D* can be determined by

$$D = \frac{4}{9\pi} \cdot \left(\frac{r}{\Delta t} \cdot \frac{\Delta E_{\text{OCP}}}{dE/d\sqrt{t}}\right)^2 \quad \text{for} \quad \Delta t \ll \frac{r^2}{D}, \quad (\text{eq. 1})$$

where *r* is the particle radius, *Δt* the pulse length, and *ΔE$_{OCP}$* the difference of the OCP due to the current pulse. The slope $dE/d\sqrt{t}$ is evaluated by a two-line fit of the pulse: the first line accounts for and subtracts the IR drop, while the second line represents the slope associated with the diffusion overpotential. [1, 9] In the technical validation chapter we briefly describe why this approach has distinct limitations.

**Analytical methods and determination of additional parameters**
Thickness of electrodes and separator is measured with a micrometer. The porosity was calculated by comparing the actual electrode density (from the electrode thickness and its weight) with the theoretical density expected from the densities of its components. Median particle sizes (D$_{50}$) of both LNMO and graphite active materials are given by the manufacturer.

The tortuosity is measured by electrochemical impedance spectroscopy (EIS) in blocking conditions according to the approach described in Landesfeind et al. [10, 11]. These EIS measurements were performed in ECC-PAT-Core-Cells (EL-CELL) using a Zahner Zennium in potentiostatic mode with an amplitude of 5 mV in the frequency range from 1 kHz to 100 kHz and 10 mHz to 50 kHz for the separator and electrode measurements, respectively. For each sample, 3 cells were run to ensure reproducibility of the measurement results. For the separators, the cells were built with copper-coated plungers to minimize the ohmic resistance. 50 µl of 1 M $LiPF_6$ in EC:EMC (3:7 w:w) were used as electrolyte. In contrast, a non-intercalating 120 µl 10 mM tetrabutylammonium perchlorate ($TBAClO_4$, Merck) in EC (Alfa Aesar):EMC (Solvionic) (3:7 w:w) electrolyte was used for the electrode characterization. Cells were built in symmetrical set-up with 120 µl of electrolyte and a glass fiber separator (Whatman GF/A). After assembly, the cells were set to rest for at least 12 h to ensure a complete wetting of the separator and electrodes.

Exchange current density *i0* of graphite is determined by EIS as well. It was conducted in galvanostatic mode with an amplitude of C/50 in three-electrode configuration vs. Li reference ring on a Zahner Zennium potentiostat in SOC increments of 5% between 0% to 100% SOC and at 2.5% and 97.5% to improve resolution on the SOC edges. The calculation of *i0* from the charge transfer resistance $R_{ct}$ is perfomed as described in more detail in [1] using the transmission line model shown in Figure 3b and the equation in Figure 3c. The numerical values plotted in Figure 3c are given in the json-file described in the data records subchapter. For LNMO, due to reproducibility issues related to the high voltages, we decide to report only a literature value for its exchange current density. For that purpose, we refer to the work from Landesfeind et al. [12], who have measured a value between 0.43-0.3 mA/cm² for a voltage of 4.4 V.

EIS measurements were analyzed with the impedance analysis software RelaxIS 3 (rhd instruments). The impedance spectra of the separators were fitted with an ECM consisting of a resistor *R* in series with a constant phase element (CPE). For the electrodes, the ECM consisted of a resistor *R* in series with a simplified transmission line model (TLM) for blocking conditions and reflecting boundary conditions with $R_{ion} \gg R_{elec}$ (see Figure 2).

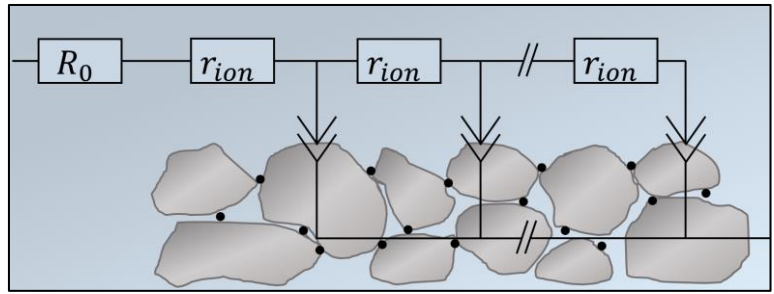

*Figure 2. ECM including a simplified TLM used for the analysis of the tortuosity measurements of the electrodes*

$$\tau = \frac{\varepsilon_e \cdot R_{ion} \cdot A \cdot \sigma}{2L} \qquad (eq.\ 1)$$

The electronic conductivity of the electrode coatings was measured by a four-point probe (Ossila). For this, the coating was delaminated from the highly conductive current collector by an adhesive tape to measure only the conductivity of the porous electrode. To ensure a good adhesion, the tape was additionally heated with a hot air blower. For each electrode, three samples were measured.

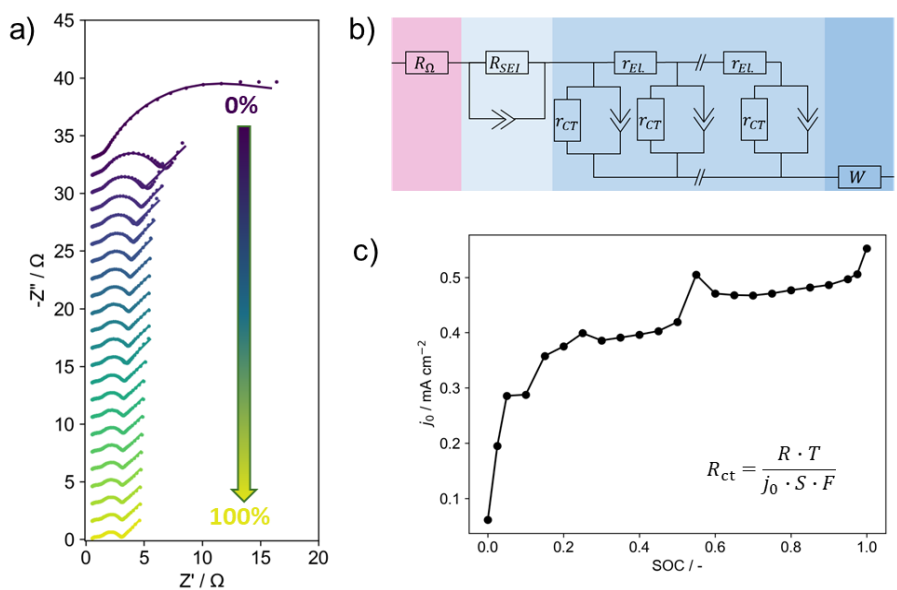

*Figure 3. Electrochemical impedance spectroscopy measurements of graphite electrodes as function of SOC (a), employed transmission line model to obtain $R_{ct}$ (b) and calculated exchange current density versus SOC (c)*

**Pouch cell manufacturing and characterization**

For pouch cell manufacturing, electrodes were punched from the electrodes described above: anodes with active surface 52x52 mm² and cathodes with active surface 50x50 mm². Electrode stacks with target capacity around 1.1 Ah were assembled in a dry room using a semi-automatic stacking equipment (MSK-111A-E, Xiamen Tmax Battery Equipments Limited). For each cell, 10 double-side positive electrode layers and 11 double-side negative electrode layers were assembled and separated with polyolefin-based microporous separator (M2000, Celgard) using the Z-folding architecture. Negative as well as positive tabs were contacted together respectively, using ultrasonic welding before integrating each stack in its pouch packaging. Pouch cells with extra pouch volume were used in this case to enable degassing after formation and to collect gas that could be generated during cycling. After a preliminary thermal sealing of the pouch edges, the cells were submitted to drying at 55 °C overnight in an oven, and then transferred into a glove-box for electrolyte activation. The same electrolyte formulation as described above was used. The activation step was performed by pipetting the required amount of electrolyte directly into the pouch cell (ca. 5.7 g); for each cell this quantity was based on the total porous volume to be filled in the cell (electrode and separator porosities) with some excess added. Finally, the last pouch edge was thermally sealed and the cells were submitted to a formation using the following protocol: C/10 CCCV charge – C/10 discharge cycle between 3.5 V and 4.9 V, at 23 °C (CV step at 4.9 V until C/50 current was reached). A good reproducibility of the formation cycle was obtained and the target capacity of 1.1 Ah was reached for all of them. The first cycle coulombic efficiency was around 91 %.

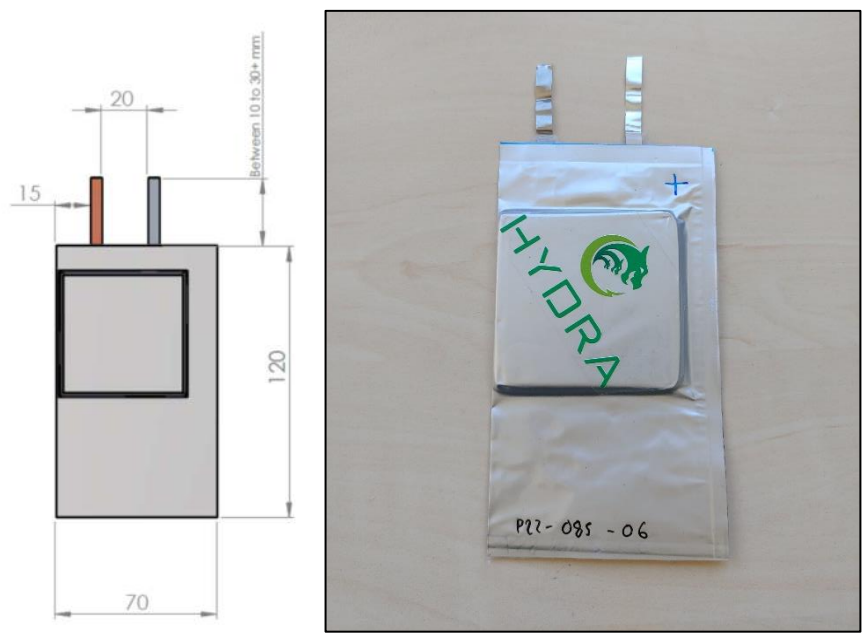

*Figure 4. Sketch and photo of the prototype 1.1 Ah pouch cell.*

**Table 2. Pouch cell design parameters.**

|  | Units | Negative electrode | Positive electrode |
|---|---|---|---|
| Electrode dimensions | mm x mm | 52 x 52 | 50 x 50 |
| No. of layers | - | 11 | 10 |
| Size of coated area | cm² | 540.8 | 500 |
| Tab length | cm | 2 | 1,9 |
| Tab width | cm | 1 | 1 |

To constrain the pouch cells with a defined pressure, a cell holder that applies a constant force according to the employed springs (spring constant $D_{spring}$=3.96 N mm$^{-1}$) was fabricated and used. To ensure a constant electrical contact of the pouch cells, the tabs were glued with a conductive silver epoxy (Chemtronics) to crocodile clamps. The nominal capacity of the cell used for the definition of the C-rates was 1 Ah.

All characterization measurements of the pouch cell at 25 °C were conducted in a Binder MK720 climate chamber. First, the pouch cell status was measured for technical validation by two consecutive charge-discharge cycles at 0.3C/0.2C between 3.5 V and 4.9 V with a constant current – constant voltage (CC-CV) charge until $I$<0.05C. In the second cycle, the cell was discharged to 50% SoC, followed by a 4 qOCV cycle with C/25 was performed. This data is included in the dataset as well. Subsequently, EIS were measured at various SOCs in 5% steps from 0% to 100% SOC and additionally at 2.5% and 97.5% SOC to have a more detailed SOC resolution for SOCs, where a large variation of the impedance is expected. EIS measurements were conducted on a Zahner Zennium in the frequency range from 5 mHz to 100 kHz and an amplitude of 20 mA. Afterwards, a qOCV cycle with C/50 was measured, followed by a rate performance test with varying discharge rates ranging from 0.05C to 2C, beginning at the lowest C-rate. During the rate performance test, the charge remained constant using a CC-CV charge protocol at 0.3C until the current dropped below 0.05C.

Then, the pouch cell was transferred into a Weiss WK340 climate chamber to perform measurements at various temperatures. First, EIS measurements at 50% SOC were measured at 5 °C, 15 °C, 25 °C, 35 °C and 45 °C after tempering the cell for 4 h at the specific temperature. After that, an additional check-up test was performed to check the influence of the temperature on the degradation of the cell. Subsequently, rate capability tests were conducted with 0.3C charge and varying discharge rates at 5 °C, 15 °C, 25 °C, and 35 °C, following the same protocol as the initial tests performed at 25 °C. As the tests at 45 °C exhibited significant capacity reduction, the data is not included herein. To minimize the impact of temperature-induced side reactions, all temperature-dependent measurements were carried out sequentially from the lowest to the highest temperature.

## Data Records

The data reported herein consists of
  i. a comprehensive parameter set describing electrode- and cell design as well as physical-chemical parameters that can serve as input for physical modelling of the lithium-ion cell,
  ii. data of various types of electrochemical measurements on electrode- and cell-level (OCV, GITT, EIS on electrode-level) that can be used to calculate thermodynamic, transport and kinetic parameters
  iii. charge/discharge measurements at varying rate for validation of electrochemical models.

The characterization results reported herein further include an array of SEM images of the pristine negative and positive electrode cross-section at different magnification to visualize active material particle morphology.

### Measured Parameters
The electrode- and cell-design parameters are reported in table 1. The microstructural parameters determined according to the procedures described in the methods section are given in table 3. Cross-sectional SEM images of the electrodes can be found on the Zenodo record.

**Table 3. Microstructural parameters.**

|  | Symbol | Unit | Negative electrode | Positive electrode | Separator |
|---|---|---|---|---|---|
| Porosity | $\varepsilon_e$ | - | 0.39 | 0.34 | 0.48 |
| Active material fraction | $a$ | - | 0.57 | 0.55 | - |
| Tortuosity | $\tau$ | - | $3.46 \pm 0.001$ | $3 \pm 0.1$ | $4.2 \pm 0.6$ |
| Particle diameter | $d$ | µm | 16.3 | 8.94 | - |
| BET surface area | $S_{BET}$ | m$^2$ g$^{-1}$ | 1.2 | 1.092 | - |
| Electronic conductivity | $\kappa_{eff}$ | S m$^{-1}$ | $62 \pm 13$ | $18 \pm 4$ | - |

### Electrochemical data on electrode-level
The electrochemical datasets generated from the graphite-LNMO 1 Ah prototype pouch cell are publicly available (doi: 10.5281/zenodo.15470746). This record contains a comprehensive collection of experimental data including GITT, quasi-OCV and rate capability tests at electrode and full-cell level, structured in Battery Data Format (BDF) [13]. The Battery Data Format (BDF) ensures that column names and units adhere to a standardized, machine-readable vocabulary, enabling maximum interoperability. The record also provides two parameter sets in JSON-LD format: (1) all parameters obtained from the measurements and (2) the parameter set with optimized parameters from the validation sequence (see 'technical validation' chapter) which are ready-to-use for the simulation (e.g. datapoints translated into functions).[14] JSON-LD as a serialization format for Resource Description Framework (RDF) provides human and machine-readable standard for expressing semantic data. The file is enriched with essential metadata descriptions of analytical methods and explicit links to the relevant datasets ensuring traceability and transparency. By using terms from BattINFO, this work creates a linked data representation of battery cells and test procedures that makes data and parameters highly reusable and interoperable across different modeling environments. As an example, our parameter set in JSON-LD can be exported using frameworks such as the Battery Parameter eXchange (BPX)[15] to other programming platforms like PyBaMM.[16] This approach not only enables seamless parameter interoperability between software tools, but also supports efficient data management and exchange while minimizing metadata loss.

The data includes GITT measurements on both electrodes versus Li reference electrode, the corresponding GITT-OCV curves and qOCV measurements obtained at C/50 for both electrodes. The data is depicted in Figure 5 and Figure 6. It should be mentioned that due to the long measurement duration and respective time at high voltage, the GITT-OCV curve of LNMO exhibits already non-negligible degradation between charge and discharging direction. It is still reported herein for the sake of completeness. For understanding rate performance of the individual electrodes and correlating this with behaviour in the pouch cell, rate capability tests on electrode-level are included as well (see Figure 7).

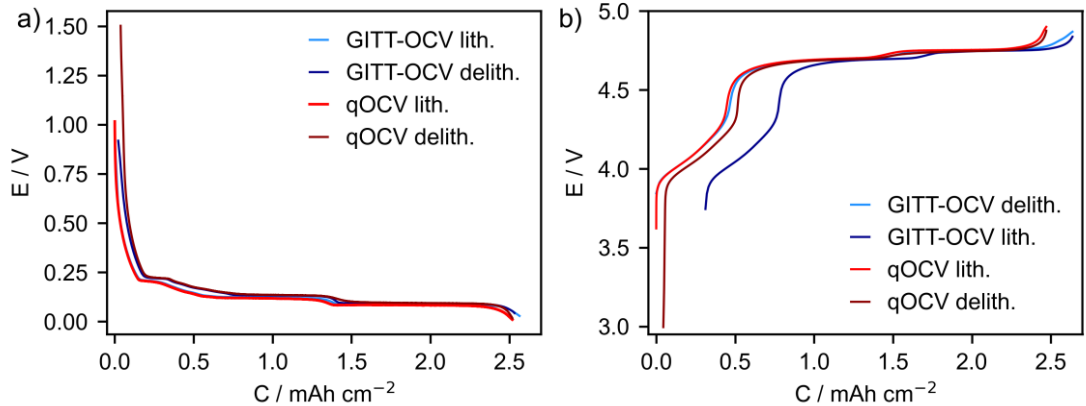

*Figure 5. OCV-curves for graphite (top) and LNMO (bottom) electrode obtained by GITT (2.5min long C/10 pulses, relaxation phase with $t < 6h$ or $\Delta U < 0.5 mV/s$) and via C/50 charge/discharge (left) and corresponding voltage vs. Li stoichiometry curve (right).*

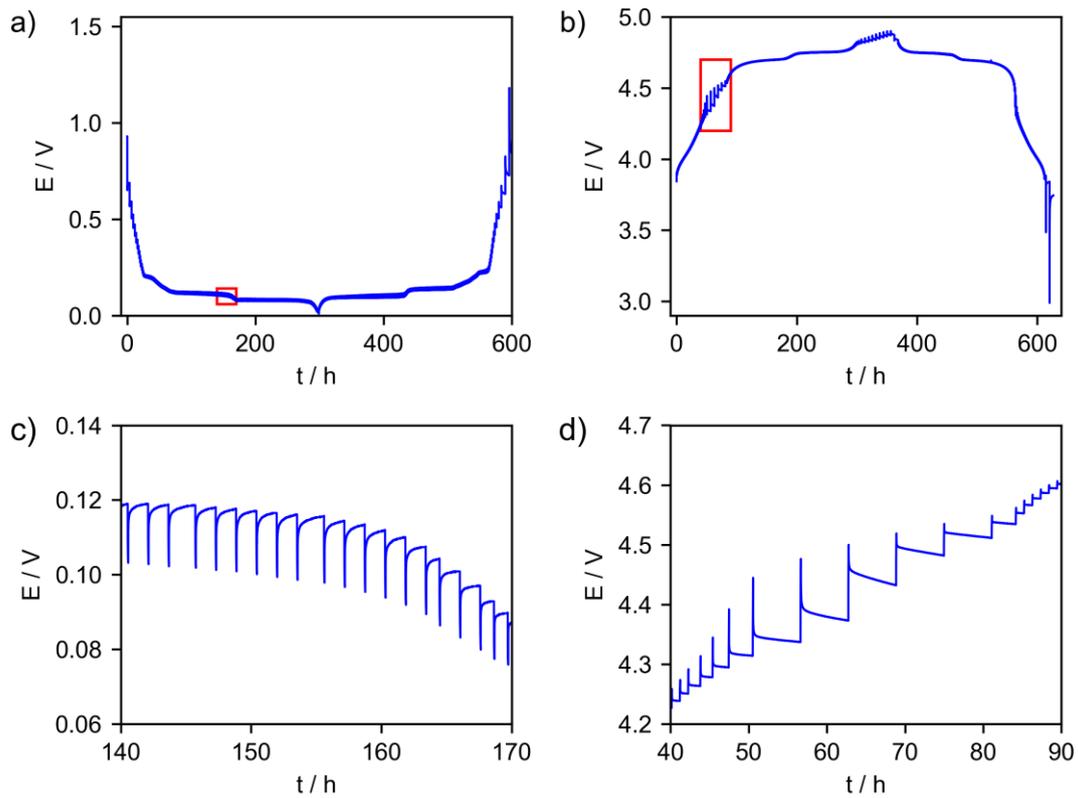

*Figure 6. GITT measurements of graphite (a,c) and LNMO ((b,d) with magnified regions in (c,d) marked with a red rectangle in (a,c)*

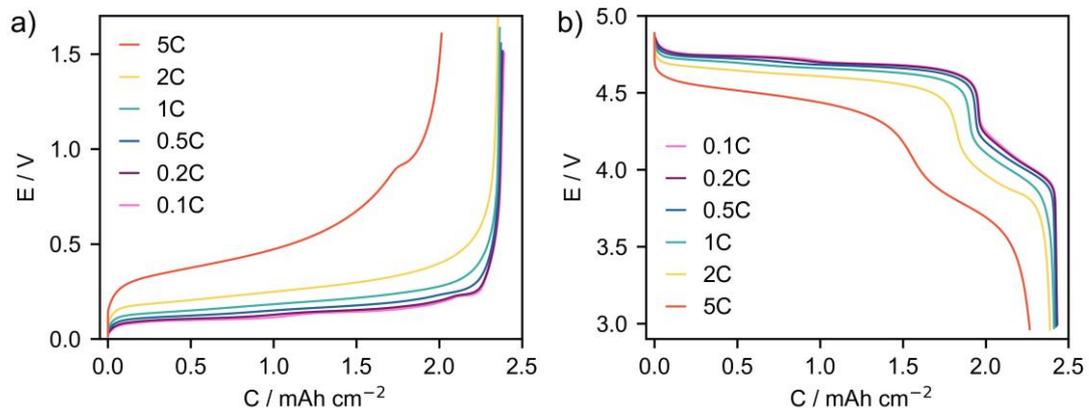

*Figure 7. Discharge curves of individual electrodes (a) graphite, b) LNMO) at varying discharge C-rates.*

**Electrochemical data on pouch cell-level**

Analogous to the electrode-level measurements, pouch cell-level data has been included in the repository to model pouch cell behaviour. The dataset includes quasi-OCV curves at 25 °C, as well as rate capability and impedance measurements at varying temperatures. Figure 8 shows the rate capability of the prototype pouch cell at varying discharge C-rates from 0.05C to 2C. As the C-rate increases, the voltage plateau decreases due to increasing overpotentials, which in turn reduces the capacity. Figure 9 highlights the effect of temperature on rate capability at 0.05C, 0.2C, 1C, and 2C. Generally, lower temperatures lead to reduced capacity at each C-rate. However, cycling at 35 °C accelerates degradation, resulting in a notably diminished capacity by the end of the rate capability test at this temperature. As shown in Figure 9, the influence of temperature on overpotentials and reduced capacity becomes more pronounced at higher C-rates. Figure 10 presents the EIS measurements of the prototype pouch cell at various SoCs and temperatures. The impedance is the highest at both low and high SoCs with a maximum at around 10% SoC. Additionally, the impedance decreases with increasing temperature.

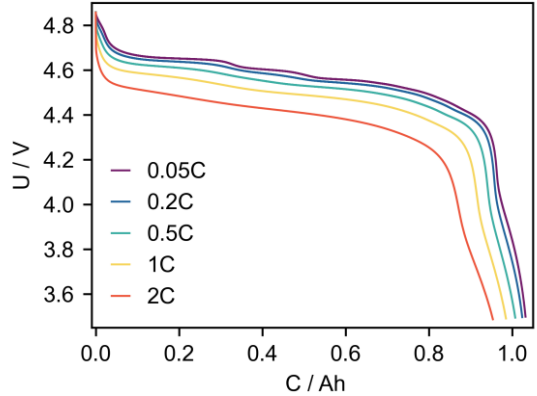

*Figure 8. Discharge curves of 1.1 Ah prototype pouch cell at varying discharge C-rate at 25 °C*

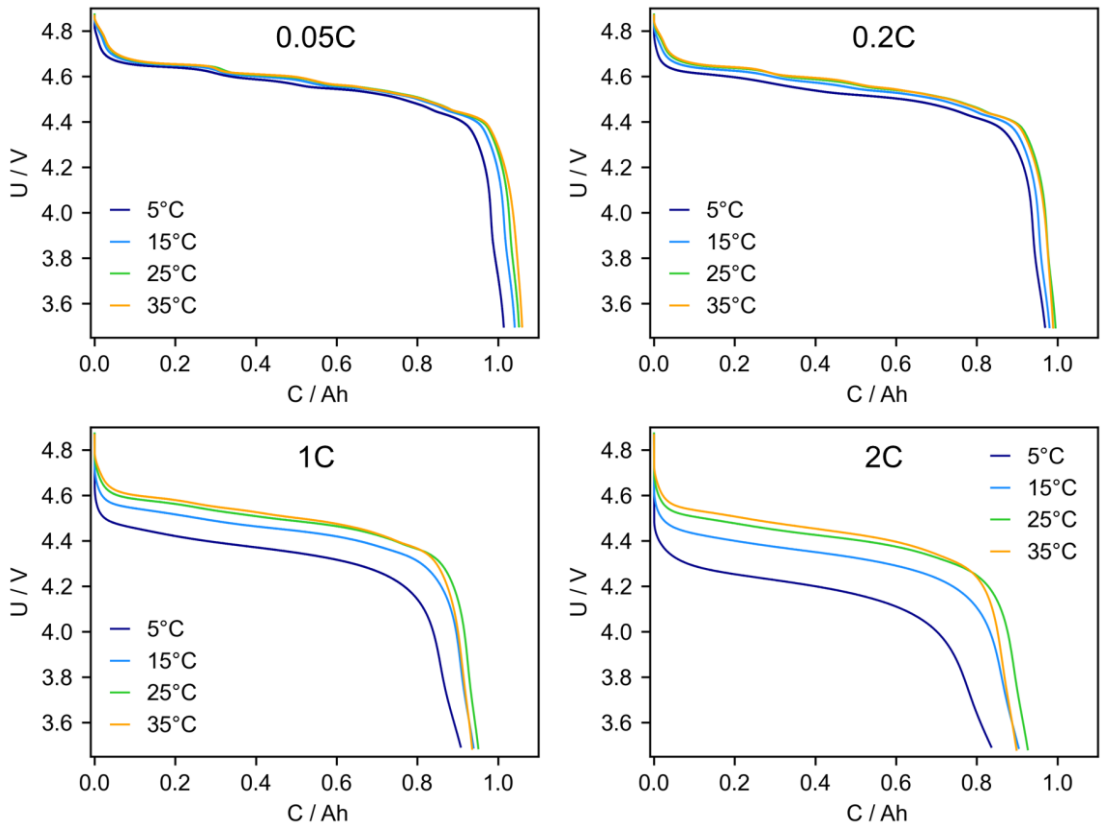

*Figure 9. Discharge curves of the prototype pouch cell at different C-rates and temperatures*

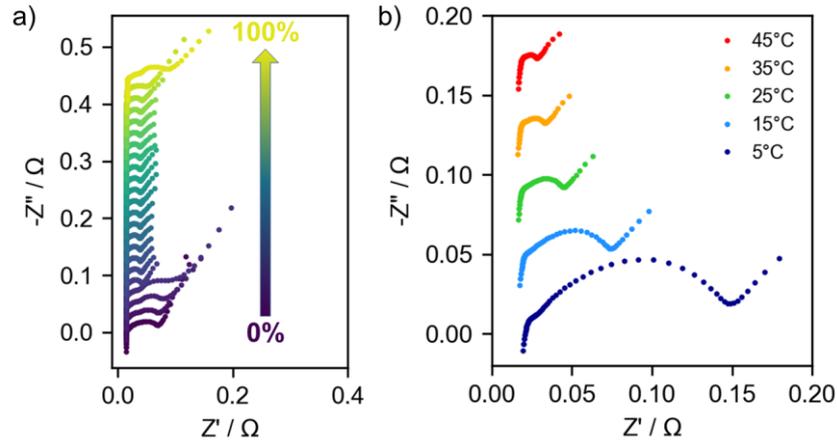

*Figure 10. EIS measurements of 1.1 Ah prototype pouch cell (a) in 5% SoC intervals (and additionally at 2.5% and 97.5% SOC) and (b) at 50% SoC (?) at varying temperatures from 5 °C to 45 °C.*

A comprehensive aging study has been conducted as well which will be disseminated in the near future.

**Technical Validation**

For the technical validation of the graphite/LNMO parameter set, the electrochemical electrode-level data reported herein was analysed to obtain the corresponding model parameters (as described in the methods section). Subsequently a full physical model was parametrized and validated versus the electrochemical measurements on pouch cell level.

Solid-state diffusion coefficients as function of SOC were estimated by analysis of the GITT measurements using the commonly employed model introduced by Weppner and Huggins [8] and a two-line fit to the experimental data. We want to emphasise that this model from 1977 uses strong simplifications that do not reflect the reality of porous electrodes with complex microstructure [17]. Furthermore, it does not apply to intercalation materials during first-order phase transitions [8, 18] evident by plateau behaviour in the (dis)charge curves of both graphite and LNMO herein. The nucleation and growth of intermediate phases, compounded by strong energy interactions between $Li^+$ and its vacancies, all contribute to the voltage relaxation response from a pulse [19]. Therefore, fitting the voltage relaxation to simply Fickian diffusion during the phase transition leads to a significant underestimations of diffusion coefficients which, in the simulation, results in unreasonably high overpotentials that stop the cell from cycling. This is in contrast to what can be observed experimentally, and clearly shows that the simulated behaviour is rather an artefact of an inaccurate measurement or simplistic model assumptions. Accordingly, for the purpose of validation of the dataset, in a first step, the SOC-dependent behaviour of the diffusion coefficients could be interpolated beyond the plateau regions where the single-phase assumption holds, aware that for an accurate description a more thorough analysis model would be required. Such a treatment has for example been done in Horner et al. using inverse modelling of the transport process [17].

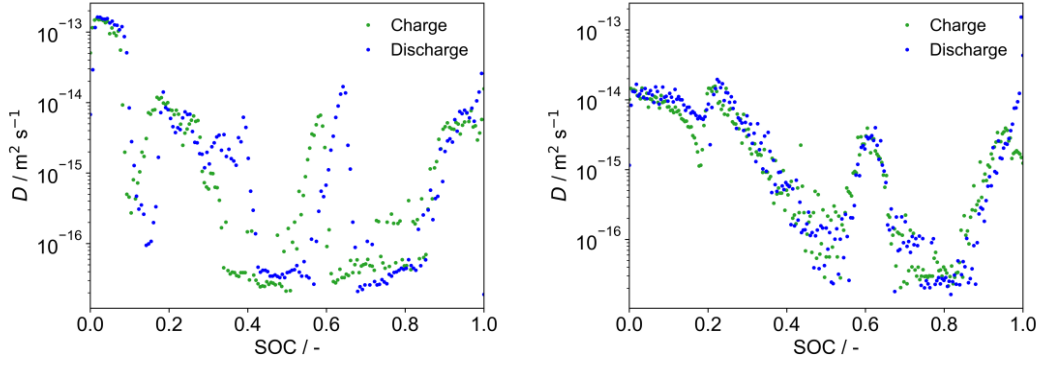

*Figure 11. Calculated solid-state diffusion coefficient for graphite (left) and LNMO (right) calculated from GITT measurements.*

The electrolyte used in the studied cell is a 1.2 M LiPF$_6$ in EC:EMC (3:7) electrolyte with 0.5% FEC. The same electrolyte without FEC (which is neglected in the following due to its low amount and function as SEI former) has been thoroughly characterized in Landesfeind et al. [20]. The relevant electrolyte characteristics for modelling are the ionic conductivity, diffusion coefficient, thermodynamic factor and transference number as function of the salt concentration. The experimental results are fitted by model equations that are given in said publication together with the corresponding fitting coefficients (in table III).

**Model calibration**

To validate our parameter set, a P2D model [21-23] simulation was conducted using the open-source software BattMo (see usage notes) based on the parameters described in this publication. As some model parameters obtained experimentally carry a distinct uncertainty, they are fitted to experimental data using the initial parameters as starting value. In addition, the model itself has intrinsic uncertainty due to simplifications and assumptions, such as the reduction of spatial and physical complexity embedded in the P2D formulation. In the calibration process, such uncertainties are compensated by adjusting parameters which capture combined effects of experimental variability and model simplifications. The calibration process herein is carried out in a two-step fashion. First, the volume fraction and the stoichiometric coefficient $\theta = c/c_{max}$ parameters are fitted to voltage curves at the lowest C-rate available (0.05C). This calibration is performed under an equilibrium assumption which is explained below. Second, transport parameters are fitted to higher-rate voltage curves (2C). In principle, there is much liberty in choosing which parameters to calibrate here; we choose for each electrode the volumetric surface area, diffusion coefficient and the Bruggeman coefficient (total 6 parameters) which are representative of interfacial kinetics, solid transport and effective electric conductivity, respectively. In addition, three Bruggeman coefficients for the electrolyte represents the ionic and electric pathways in the electrodes as well as the separator (3 parameters). In total there are 9 parameters that are calibrated. The volumetric surface area SA/V is calculated from the experimental parameters according to $SA/V = S_{BET}\, \rho_{AM}\, a$, where $S_{BET}$ is the surface area measured by N$_2$ physisorption, $\rho_{AM}$ the density of the active material, and $a$ the volume fraction of the active material. Since the Bruggeman coefficient is a model parameter that is usually inferred from effective properties, not the other way around, we choose to report tortuosities in the following. From the Bruggeman coefficients $b$, we calculate the tortuosity $\tau$ as $\tau = \varepsilon^{-b}$, see Landesfeind et al (cite the Landesfeind tortuosity paper). The effective electronic conductivity is calculated as $\kappa_{eff} = \varepsilon^{-b}\kappa$.

For the equilibrium assumption we assume instantaneous diffusion and transport processes. This means that the stoichiometric coefficient can be described by

$$\theta_e(t) = \theta_{e,100} + sgn_e \frac{I}{FV_e \varepsilon_e a_e c_{e,max}}(t - t_0) \quad \text{(eq. 2)}$$

for each electrode $e$. Here $F$ is the Faraday constant, $V$ the total volume, $\varepsilon$ the volume fraction, and $\theta_{100}$ the stoichiometric coefficient at 100% SOC of the full cell. The sign is positive for the positive electrode (PE) and negative for the negative electrode (NE). Assuming low rates, the cell voltage is given by

$$U(t) = OCP_{PE}(\theta_{PE}(t)) - OCP_{NE}(\theta_{NE}(t)) \qquad (eq.\ 3)$$

For the calibration, we choose $\theta_{100}$ and the product $m = V\ \varepsilon\ a\ c_{max}$ as adjustable parameters. Since we trust the volume, active material fraction (given by the manufacturer) and $c_{max}$, this is equivalent to calibrating the volume fraction or the usable electrode capacity. Mathematically, the calibration is solved by minimizing the root mean square (RMS) difference between $U$ and $U_{exp}$. Here, we solve this with a gradient-based optimization method, namely the limited-memory Broyden-Fletcher-Goldfarb-Shanno (L-BFGS) algorithm, cf [24]. The gradients are obtained automatically due to the automatic differentiation capabilities of BattMo.

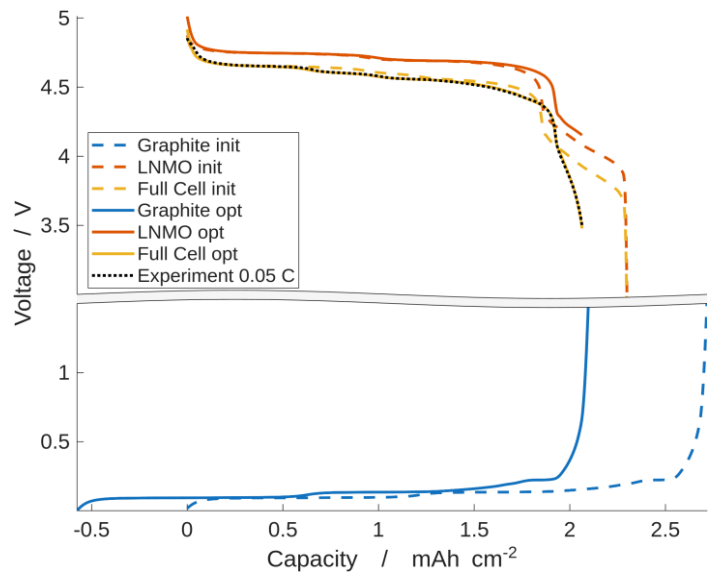

Figure 12. Cell balancing under equilibrium assumption

**Table 4.** *Results before and after equilibrium calibration.*

|     | $\theta_{100}$ initial | $\theta_{100}$ calibrated | $\varepsilon$ initial | $\varepsilon$ calibrated |
|-----|------------------------|---------------------------|------------------------|---------------------------|
| NE  | 1.0                    | 0.7859                    | 0.61                   | 0.5993                    |
| PE  | 0.14                   | 0.1282                    | 0.66                   | 0.6733                    |

The high-rate calibration is also performed by minimizing the same RMS difference objective function, but now under the constraint that $U(t)$ is not given by equation 3, but rather by the full electrochemical dynamics of the P2D model. Solving such systems is costly: it is nonlinear, time and space-dependent and can contain many degrees of freedom. Therefore, it is important to reduce the number of P2D model evaluation during the optimization, making a gradient-based optimization method such as L-BFGS suitable.[24] With the use of the adjoint problem, the cost of obtaining the gradients is independent of the number of parameters.

**Table 5.** *Results before and after high-rate calibration*
(PE/NE = positive/negative electrode, S = separator)

|    | *volumetric surface area,* $SA/V$ / $m^{-1}$ | | *solid-state diffusion coefficient* $D_s$ / $m^2\ s^{-1}$ | | *tortuosity* $\tau$ / - | | *Electronic conductivity* $\kappa_{eff}$ / $S\ m^{-1}$ | |
|----|--------|-----------|---------------|------------|---------|------------|---------|------------|
|    | initial | calibrated | initial | calibrated | Initial | calibrated | initial | calibrated |
| NE | 1.54E6 | 4.90E6 | 1E-13 / 1E-14 | 1.24E-13 | 3.46 | 2.18 | 62 | 21 |
| PE | 2.52E6 | 1.01E5 | 1E-14 | 2.82E-14 | 3 | 2.89 | 18 | 5 |
| S  | - | - | - | - | 4.2 | 2.90 | - | - |

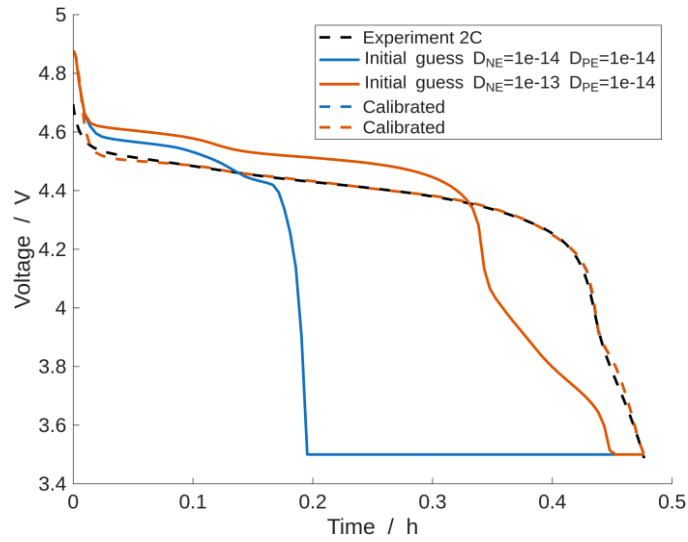

*Figure 13. Results after high-rate calibration at 2C using two different initial values for the diffusion coefficient of the negative electrode. The calibrated curves lie on top of each other.*

**Validation**

The calibrated values under equilibrium assumption (0.05C) are shown in Table 4 and the resulting voltages are seen in
Figure 12. As one can see, the calibrated volume fractions only change slightly (< 2%) versus the initial values. This mean the parameters describing the capacities were determined with good accuracy. The shift of the negative electrode OCP in
Figure 12 is a consequence of calibrating for the actual balancing of the electrodes in the full cell and the corresponding *effective* N/P ratio. The cell balancing takes into account the loss of lithium inventory during the formation cycles and additional aging between time of production and testing (in our case 14%) which shifts the two electrode OCP curves with respect to each other.

The rate-dependent parameters obtained from the second stage calibration at 2C are shown in Table 5. In Figure 13 we see the resulting voltage. The initial guess (blue) is from using the data from the equilibrium calibration. The calibrated voltage (red) shows some discrepancy from the experimental voltage in the beginning, but this is believed to be due to the use of a 1D geometry rather than a full 3D geometry. Figure 14 shows the experimental voltages, and the voltages obtained from the fully calibrated P2D model over the available experimental C-rates. The match is good, but best for the lowest and highest rates, since we have tuned the parameters for these cases.

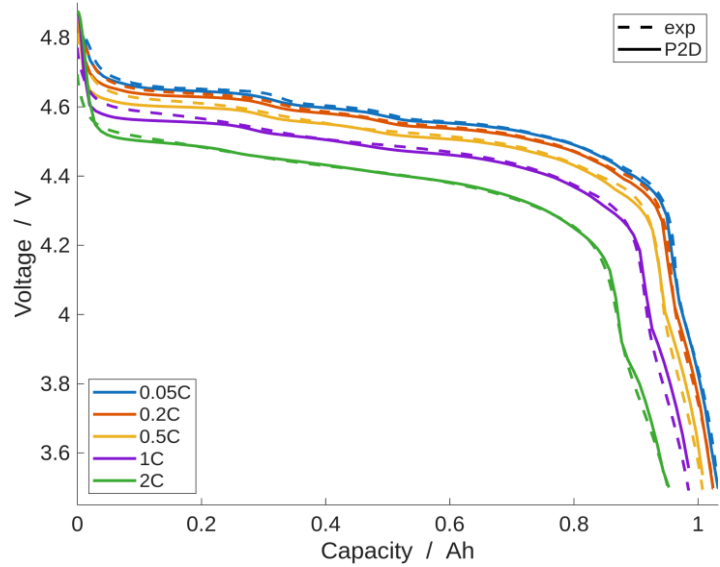

*Figure 14. Experimental voltages and P2D model results over different discharge rates.*

Commenting on the results of the parameter calibration, the differences between measured and calibrated parameters for the high-rate calibration need to be interpreted with the DFN model structure in mind. As described above, the calibrated parameters were chosen to represent the individual processes in the DFN model which means that during calibration, the optimization algorithm redistributes the mismatch between experiment and model across this set of parameters and over the parameters within each term it contributes to. This implies that as several parameters affect the same observables, the solution is not unique and that due to grouping with other parameters they are not individually identifiable. For example, the same change of the solid-state diffusion can be accommodated by changing *Ds* or particle radius.

Nevertheless, the parameter changes reported herein are physically reasonable and fall within uncertainties associated with both the experiments and the model to a realistic degree. This suggests that both the experiments provided good estimates as model input and that the model calibration described herein yields a meaningful set of effective parameters. In general, uncertainty from the model can be ascribed to several model assumptions: 3D effects are not captured in the P2D model, the use of uniform, spherical parameters, constant parameters for parameters which are normally a function of SOC (*Ds* and $i_0$), no double layer capacitance and no thermal model (all parameters are isothermal at 25 °C) included which especially at the higher c-rates is a very strong assumption. In the following, we assess the parameter changes during the calibration in more detail.

- Among all parameters, the volumetric surface area shows the largest deviation between initial and calibrated value. It enters the DFN model through its scaling effect on the exchange current density and is further entangled with other microstructural parameters. The rather large change of the parameter values for both electrodes can be well ascribed to a combination of the intrinsic limitations of the BET surface area as measurable quantity to capture the relevant electrochemically active surface area [25] and the fact that the kinetic parameters were taken from literature and not measured for the specific materials at hand.
- The limitations of Ds derived from the standard Weppner and Huggins method has been elaborated above. Despite the very strong assumptions and simplification, the final values are surprisingly close to the initial ones. This shows that well measured and analysed GITT measurements can still serve as reasonable baseline when advanced approaches like inverse modelling are not available. To address the fact that for a large part of the SOC, the Ds for both graphite and LNMO is not accurately captured by the above method, the use of a single SOC-independent value is justified.
- Furthermore, all three tortuosity values were corrected downwards to obtain a good model fit which likely reflects uncertainties in the determination of the porosity which is used in its calculation and, again given the model structure, the intrinsic difficulties in accurate determination of electrolyte transport parameters (as recently described by Lehnert et al. [26]).
- Similarly, the effective electronic conductivity which is corrected to lower values cannot be uniquely separated from uncertainties in the kinetic parameters (not measured for the specific materials) and the surface area, thus, its calibrated value likewise represents an effective parameter that likely compensates for these coupled influences.

Ultimately, the calibrated values serve as a consistent set of effective parameters that allows the DFN model to reproduce the observed cell behavior within the uncertainty of both the experiments *and* the model.

## Usage Notes

This parameter set and validation data is made available to the community to parameterize different types of simulation models and to test algorithms for parameter extraction similar to what has been done as part of the technical validation reported within this contribution.

All the numerical studies in this paper are performed using BattMo, a free, open-source software for simulation of electrochemical systems. It has primarily been developed for DFN-models of Li-ion batteries in 1, 2 and 3D, but has also been applied to other battery chemistries and electrolyzers. BattMo is written in MATLAB and is available at https://github.com/BattMoTeam/BattMo. A Julia version is in active development and available at https://github.com/BattMoTeam/BattMo.jl. The data and the code for the simulations used in this paper can be found in the repository https://github.com/BattMoTeam/2024-Hydra-calibration-paper.

## Code Availability

All scripts for reproducing the figures in this paper are available in the repository https://github.com/BattMoTeam/2024-Hydra-calibration-paper under a GPL 3.0 license. For the results here, MATLAB R2025a has been used.

## Acknowledgements

The authors gratefully acknowledge funding from the European Union's Horizon 2020 research and innovation programme under grant agreement no. 875527 (HYDRA) and Horizon Europe grant no. 101103997 (DigiBatt). The authors would like to thank the partners of the both consortia for their contributions and the industry partners for providing the materials.

## Author contributions

- Christina Schmitt: experiments, analysis, conceptualization, writing
- August Johansson: numerical methodology and software development, simulation, writing
- Xavier Raynaud: numerical methodology and software development, simulation
- Eibar Joel Flores Cedeño: supervision of modelling work, conceptualization, writing
- John Mugisa: data management, writing

- Dane Sotta: cell manufacturing, writing
- Agathe Martin: cell manufacturing
- Nicolas Schaeffer: cell manufacturing
- Cédric Debruyne: cell manufacturing
- Yvan Reynier: cell manufacturing
- Simon Clark: supervision of modelling work, conceptualization, writing
- Dennis Kopljar: supervision of experimental work, conceptualization, writing